\documentclass[12pt,preprint]{aastex}

\usepackage[english]{babel}
\usepackage[utf8x]{inputenc}
\usepackage[T1]{fontenc}

\usepackage{amsmath}
\usepackage{amssymb}
\usepackage{graphicx}
\usepackage[colorlinks=true, allcolors=blue]{hyperref}
\usepackage{booktabs}
\usepackage{ulem}
\usepackage{lineno}

\shorttitle{}
\shortauthors{Haghighipour \& Maindl}

\begin{document}

\title{Building Terrestrial Planets: \\
\normalsize Why results of perfect-merging simulations are not quantitatively reliable 
approximations to accurate modeling of terrestrial planet formation}

\author{Nader Haghighipour\altaffilmark{1,2,3} and Thomas I. Maindl\altaffilmark{4,5}}

\altaffiltext{1}{Planetary Science Institute, Tucson, AZ, USA}
\altaffiltext{2}{Institute for Astronomy, University of Hawaii-Manoa, Honolulu, HI, USA}
\altaffiltext{3}{Institute for Advanced Planetary Astrophysics, Honolulu, HI, USA}
\altaffiltext{4}{SDB Science-driven Business Ltd, Larnaca, Cyprus} 
\altaffiltext{5}{Department of Astrophysics, University of Vienna, Vienna, Austria}
\email{nader@psi.edu}

\begin{abstract}
 Although it is accepted that perfect-merging is not a realistic outcome of collisions, some researchers state
 that perfect-merging simulations can still be considered as quantitatively reliable representations of the final
 stage of terrestrial planet formation. Citing the work of Kokubo \& Genda [ApJL, 714L, 21], they argue that the
 differences between the final planets in simulations with perfect-merging and those where collisions are resolved
 accurately are small, and it is, therefore, justified to use perfect-merging results as an acceptable approximation to realistic
 simulations. In this paper, we show that this argument does not stand. We demonstrate that when the mass lost
 during collisions is taken into account, the final masses of the planets will be so different from those obtained
 from perfect-merging that the latter cannot be used as a valid approximation. We carried out a large number of SPH
 simulations of embryo-embryo collisions and determined the amount of the mass and water lost in each impact.
 We applied the results to collisions in a typical perfect-merging simulation and showed that even when the mass-loss
 in each collision is as small as 10\%, perfect-merging can, on average, overestimate the masses of the final
 planets by $\sim 35\%$ and their water-contents by more than 18\%. Our analysis demonstrates that, while perfect-merging
 simulations are still a powerful tool in proving concepts, they cannot be used to make predictions, draw quantitative
 conclusions (especially about the past history of a planetary system) and serve as a valid approximation to, or in lieu of
 the simulations in which collisions are resolved accurately.
\end{abstract}

\keywords{Planet formation(1241), Solar system formation(1530), N-body simulations(1083)}

\section{Introduction}

The classical model of terrestrial planet formation has had significant contributions 
to our understanding of the formation of Earth and the origin of its water. This model 
that uses N-body integrations to simulate the collisions and orbital evolution of 
gravitationally interacting bodies has demonstrated that terrestrial planets form 
through giant impacts among planetary embryos 
\citep[e.g.,][]{Kokubo98,Chambers98,Chambers01b,Agnor99,Chambers01a,Kokubo06}, proved that 
Earth’s water was brought to its accretion zone by hydrated protoplanetary bodies from the 
outer regions of the asteroid belt \citep[e.g.,][]{Morbidelli00,Obrien14,Obrien18,Haghighipour16},
and presented models to account for the small mass of Mars 
\citep{Walsh11,Izidoro14,Clement19a,Nesvorny21} and even a possible origin for 
Mercury \citep{Clement19b,Clement21a,Clement21b,Clement21c}.

As successful as it has been, the classical model and its N-body integrations suffer 
from an important shortcoming: they consider collisions to be perfectly inelastic. 
That means, when two objects collide, the colliding bodies are completely merged and 
no breakage, fragmentation, and debris is considered. 
This perfect-merging recipe also assumes that during a collision, 
the internal compositions (e.g., water-contents) of the colliding bodies 
stay intact and when collided, the entire composition of each object is transferred to 
the merged body. As a result, planets 
that are produced in the classical simulations carry more mass and water than they 
would have if simulations had included fragmentation, and had the transfer of water 
been treated properly. As the overestimation of mass directly affects the dynamical
evolution and orbital architecture of the final planets, when combined with the 
overabundance of water, it indicates that planets produced by the classical model 
cannot be considered as reliable representations of observational and geological data. 
In other words, the classical model cannot be quantitative and cannot make predictions.

In reality, collisions are not perfectly inelastic. They produce fragments 
and large amounts of debris. A realistic model of terrestrial planet formation needs 
to take this into account. 
In the past decade, several groups have addressed this issue by resolving collisions using
high-resolution simulations
\citep[e.g.,][]{Marcus09,Marcus10,Kokubo10,Stewart12,Leinhardt12,Chambers13,Maindl13,Maindl14,Carter15,
Carter18,Bonsor15,Wallace17,Burger18,Burger20}. However, despite its fundamental shortcomings, 
the classical model and its perfect-merging approach have stayed deeply popular. 
This popularity, that is mainly because of the ease of computations, 
is primarily based on the results of the fundamental works of \citet{Kokubo10} and \citet{Genda12}. 
These authors carried out a large number of Smooth Particle Hydrodynamics (SPH) simulations 
of the collisions of planetary 
embryos and, following the seminal work of \citet{Agnor04}, identified an empirical value 
for the impact velocity that they used as the criteria to determine the outcome of embryo-embryo collisions.
\citet{Kokubo10} showed that for the values of the impact velocity above this threshold, the two 
colliding bodies may break, shatter, or undergo a hit-and-run event \citep[similar results had also 
been obtained by][]{Asphaug09}. However, for lower values, collisions may produce one large body along with
debris and possibly even small fragments. These authors argued that because during the formation of (terrestrial)
planets, it is the large body that has the most contribution, it is justified to
ignore the debris and few fragments, and consider/define such events as accretion (or merging). 
\citet{Kokubo10} showed that in their SPH simulations to which their above definition of accretion applies, 
the final mass and number of terrestrial planets 
were similar to those obtained from N-body integrations with perfect-merging. 
The supporters of the perfect-merging scenario argue that, although the accretion events by 
\citet{Kokubo10} are not 100\%, the difference between their outcomes and those of 
perfect-merging simulations is not large enough, and results of perfect-merging simulations can
be used as a valid and reliable approximation to simulations where collisions are resolved accurately.

Although the conclusions drawn by \citet{Kokubo10} and \citet{Genda12} are valid within the context of 
their simulations, when used as the basis for the argument that the error in using
perfect-merging recipe is negligible, do not provide a solid ground. The reason lies in
the fact that when labeling a collision as accretion, the amount of the mass that did not contribute
to the formation of the single body (e.g., debris) was ignored. \citet{Kokubo10} and \citet{Genda12} 
argued that because their focus was on modeling the formation of terrestrial planets, 
only collisions that produced a single body were the most relevant ones, and for that reason, they
categorized collisions into only two categories of hit-and-run (non-merging) and non-hit-and-run 
(merging) events. As a result, many cases in which debris and/or some fragments were produced and 
scattered out of the accretion zone were labeled as accretion events without  
their mass-losses being taken into account. 
\citep[We would like to note that in a later study][returned to their previous N-body simulations and
calculated the debris produced in each system. We will dicsuss their results in more detail in
sections 3.4]{Genda15}.

While it is possible that in systems where the final planet forms after only one or two collisions, 
the total mass-loss may be small, when considered in the larger context of planet formation, where embryos are
subject to many collisions, the cumulative loss of mass (and volatiles such as water) may not be negligible. 
The purpose of this paper is to demonstrate the latter. We have carried out a large number of SPH simulations
of embryo-embryo collisions and calculated the amount of the mass and water lost in each case. We have also 
carried out a suite of simulations of terrestrial planet formation using the classical model, and determined 
the mass and water-contents of their final planets using the perfect-merging approach. 
We will show that when the mass-loss obtained from SPH simulations is applied to the simulations of
classical model, the cumulative loss of mass and water during the formation of a planet
will be so large that the final outcome of the perfect-merging scenario will be quantitatively unreliable. 

The outline of this paper is as follows. 
In the next section, we will explain the details of our methodology and approach. In Section 3, 
we analyze the results and compare them with previous studies. In Section 4, we discuss the implications
of our findings and present our concluding remarks.

\section{Numerical Simulations}

\subsection{Perfect-Merging (Classical Model)}
We began by carrying out perfect-merging simulations in the context of the classical model. 
Each simulation started with a disk of protoplanetary bodies extending from 0.5 AU to 4 AU. The disk
consisted of 1000-1200 planetesimals and 76 planetary embryos. The bi-modality of the disk 
was to include dynamical friction \citep[e.g.,][]{Ida92,Obrien06,Morishima08} 
and to ensure that the starting systems would 
be consistent with the outcomes of Runaway and Oligarchic growths \citep{Kokubo98,Kokubo00,Darriba21}. 
The embryos were distributed according to the surface density 
$10 \, ({\rm g \, cm^{-2}})\,{(r/{\rm AU})^{-\alpha}},\, \alpha=0.5, 1, 1.5$,
and planetesimals were distributed uniformly. The masses of embryos,
ranging between 0.015 and 0.071 Earth-masses, were scaled with their distances to the Sun 
$(r)$ and the number of their mutual Hill radii $(∆)$ as  ${r^{3(2-\alpha)/2}} {∆^{3/2}}$  
\citep{Kokubo00,Raymond05a,Raymond05b,Raymond09,Izidoro13,Izidoro14,Haghighipour16}.
The distances between planetary embryos were randomly chosen to vary between 5 and 10 mutual Hill radii.

We considered every planetesimal to have an initial mass of 20\% lunar-mass. As explained in section 3.2,
we did not treat planetesimals as test particles and included their masses when calculating the final mass
of a planet. We assigned water to all bodies according to the following: Bodies interior to 2 AU 
were considered to have a water-mass fraction 
of $10^{-5}$, those between 2 AU and 2.5 AU to have 0.1\% water in mass and those beyond 2.5 AU to have a 
water-mass fraction of 10\%. We also assumed that no earlier radial mixing had smoothed out the water 
distribution in the protoplanetary disk \citep[e.g.,][]{Carter15}. 

Simulations were carried out both with and without giant planets. When including giant planets, two cases of
Jupiter-only and Jupiter-Saturn were considered. We placed these planets in their current semimajor axes 
and carried out simulations for the values of their eccentricities equal to 0, 0.1, and their current values.

In total, we carried out 42 simulations. We integrated each system for 200 Myr using the hybrid routine 
in the N-Body integration package MERCURY \citep{Chambers99}. The time-step of integrations were set to 
6 days. In general, in each simulation, 1 to 4 planets formed interior to 2.1 AU with most simulations forming 3 planets.
A typical simulation produced 45 to 60 embryo-embryo collisions (giant impacts) and 
200 to 320 embryo-planetesimal collisions. The number of giant impacts that resulted in the formation of a
planet varied between 6 and 11. Collisions among planetesimals were not included. 

Figure 1 shows a sample of our results. In this simulation $\alpha=1.5$,
Jupiter and Saturn are in their current orbits and carry their current eccentricities. The protoplanetary
disk in this simulation has a total mass of $\sim$5 Earth-masses with a planetesimal to embryo mass-ratio
of $\sim 1$. The color-coding demonstrates the water-mass fraction (WMF) of each object.
We will use this simulation and its final results (shown by the panel at 200 Myr) to demonstrate the errors in
using perfect merging as a quantitative model. We would like to note that the differences between
the orbital assembly as well as the physical and compositional properties of the final planets in figure 1 and those 
of the terrestrial planets in our solar system are expected results that have roots in the fact that all 
simulations of planet formation are stochastic and their results must be studied statistically.

\subsection{SPH Simulations of Embryo-Embryo Collisions}

To determine the amount of mass and volatiles (specifically water) lost during an impact,
we carried out a large number of SPH simulations of embryo-embryo collisions
using our 3D, solid-body, continuum-mechanics SPH code MILUPHCUDA \citep{Maindl13,Schafer16,Schafer20}.  
Our code solves the continuity equation and the equation of the conservation of momentum in 
continuum mechanics, and includes self-gravity. We  discretized the solid body continuum into 
mass packages (i.e., SPH particles) and used the locations of these particles as the sampling points of the 
numerical method. The SPH particles move similar to point masses following the equation of motion. 
Each particle carries all physical properties of the part of the solid body that it represents (e.g., mass, 
momentum, and energy). To simulate the plastic behavior of solid materials, we followed \citet{Collins04} 
and \citet{Jutzi15}, and used pressure-dependent strength models. Fractures and brittle failures were handled by 
implementing the Grady-Kipp fragmentation model \citep{Grady80,Benz95}. We ensured first-order 
consistency by applying a tensorial correction as presented by \citet{Schafer07,Schafer16}. Dissipation of kinetic 
energy into heat was modeled by tracking inner energy including viscous energy terms originating from 
artificial viscosity \citep{Monaghan83}. 

We considered embryos to be silicate rocks composed of pure basalt or basalt and water. 
Basaltic rock is commonly used as the material of rocky bodies from centimeter-sized grains to the mantles of large
asteroids such as Ceres \citep{Melosh97,Agnor04,Michel09,Nakamura09}. We used the Tillotson equation of state 
\citep{Tillotson62,Melosh96} to model the materials of the colliding bodies. The parameters for basalt and 
water-ice were taken from \citet{Melosh96} and \citet{Benz99} and are given in Table 1. We also implemented 
a damage model using 
Weibull distribution of strain level $\epsilon$ given by $n(\epsilon)=k{\epsilon^m}$ \citep{Weibull39}.
We used the Weibull parameters $(m,k)=(16,{10^{61}}\,{\rm {m^{-3}}})$ for basalt \citep{Nakamura07} and 
$(9.1,{10^{46}}\,{\rm {m^{-3}}})$ for ice \citep{Lange84}, respectively. 

Figure 2 shows the geometry of an impact. We considered the target to be at rest and allowed the impactor to collide 
with it while moving with the impact velocity $v_{\rm {imp}}$ given in the units of mutual escape velocity
$v_{\rm esc}^2=2G({m_{\rm t}}+{m_{\rm i}})/({R_{\rm t}}+{R_{\rm i}})$. Here, $m_{\rm t}$ and $m_{\rm i}$ are the masses
of the target and impactor, and $R_{\rm t}$ and $R_{\rm i}$ represent their corresponding radii, respectively. 
The impact angle was chosen such that $\alpha=0$ would be a head-on collision and $\alpha={90^\circ}$ would
represent a grazing impact. We chose the impactor to be 100\% basaltic rock and the target to be a silicate rock 
composed of basalt with 30\% water in its mantle and crust. When modeling collisions, we considered water to
be surface ice on the top of a solid layer (a water shell on the top of the mantle).

Figure 3 shows a sample of the results. Here we show the amount of water lost in an impact between two Ceres-sized
bodies for different values of their impact angle and impact velocity. The color-coding represents water-loss per 
collision in terms of the total mass of the colliding bodies 
[wt-\% short for weight-percent, representing mass-fraction]. 
 We define water-loss as the fraction of total water that is no longer gravitationally bound to the 
``surviving bodies’’ (i.e., the outcome of an impact). For a given material strength, the surviving bodies 
depend on the impact velocity and angle. In the simulations of figure 3, three types of outcome were obtained:
merging, where one major body was produced along with debris that was lost; hit-and-run, where two major bodies 
survived along with debris that was lost; and erosion, where we considered only the two largest fragments as 
surviving bodies \citep[for a quantitative definition of these outcomes, we refer the reader to][]{Leinhardt12}. 
We would like to emphasize that the collision outcome ``merging’’ does not refer to ``perfect merging’’
in which the two colliding bodies undergo perfectly inelastic collision and are completely merged. 
Unlike in the perfect-merging scenario, a merging event produces debris that in most cases is lost to space.

Because we are interested in both the loss of mass and water, we chose to show the amount of water-loss as 
it can also be used as a proxy for mass-loss. As shown here, the loss of water (and mass) varies from 
very small values close to 10\% \citep[in agreement with the results of][]{Kokubo10} 
to more than 90\%. For example, 
42\;\% of our collisions resulted in merging with an average water-loss of 7\;wt-\% (population variance = 1\;wt-\%), 
33\;\% resulted in hit-and-run with an average water-loss of 5\;wt-\% (population variance = 0.2 wt-\%), 
and 25\;\% resulted in erosion with an average water-loss of 74\;wt-\% (population variance = 4 wt-\%).
Figure 3 also shows that the loss is small for low impact velocities irrespective of the impact angle, 
and for collisions close to grazing impacts. 

It is important to note that the overall average of water-loss in all our simulations is $\sim$ 23 wt-\%.
However, this value has no physical meaning because the initial values of impact angles and velocities in our 
simulations had not been chosen to resemble their distribution in a typical planet formation scenario.
They had been selected to cover a large region of the parameter-space. 
In the next section, we use results shown in figure 3 to calculate the total loss in  mass and water
for the final planets of figure 1.

\section{Results and Analysis}

\subsection{Calculation of Mass- and Water-Loss}

In this section, we will demonstrate that when the results of the SPH simulations of section 2.2
are applied to the perfect-merging planets of section 2.1, the cumulative loss of mass and water 
will not be negligible.

We start by using a simple system and some simple assumptions to portray the general picture of the concept 
that we are trying to show (i.e., the significance of mass-loss in multiple collisions).
Let's assume, for the mere sake of argument, that the definition of accretion as stated by 
\citet{Kokubo10} and \citet{Genda12} stands. That is, if two embryos collide, regardless of how much debris
and fragments are produce, as long as one main body is formed, that event is labeled accretion 
(or merging). Also, merely for the purpose of demonstration, let's assume that in any embryo-embryo 
collision, a fixed fraction $\eta$ of the total colliding masses is lost after collision. That is, if an embryo 
with a mass $m_1$ impacts an embryo with a mass $m$, irrespective of their material strength and their angle 
and velocity of impact, the post-impact mass of the merged body will be $(1-\eta)(m+{m_1})$ indicating that
$\eta(m+{m_1})$ amount of mass was lost due to the collision. This assumption is not fully realistic because the 
mass of the collisional debris varies based on the mass, material strength, and dynamical properties of the 
colliding bodies. However, for the purpose of demonstrating the concept, it is useful.

Let's now assume that the first merged body with the mass $(1-\eta)(m+{m_1})$ is impacted by an 
embryo with a mass $m_2$. According to our assumption, the amount of mass lost during this impact 
will be $\eta[(1-\eta)(m+{m_1})+{m_2}]$ and the new merged body will have a total mass of 
$(1-\eta)[(1-\eta)(m+{m_1})+{m_2}]$.
Continuing in the same fashion, the total amount of mass lost after $n$ impacts will be equal to

\begin{equation}
{\rm Total}\,\,{\rm Mass}\,\,{\rm Loss}=\sum_{j=1}^n\eta{(1-\eta)^{n-j}}\,m\,+
\sum_{j=1,...,k}^n\eta{(1-\eta)^{n-k}}\,{m_k} \qquad k=1,2,...,n \,.
\end{equation}

\noindent
In this equation, $m_k$ is the mass of the embryo that generates the $k$-th impact. It is important to note that
the summation over $j$ in the second term is carried out for each value of $k$, separately. 

A closer look at the summation terms in equation (1) indicates that these terms are in fact
geometric series and can be calculated analytically. For instance, the first summation, corresponding to
the contribution of embryo $m$ to the total mass-loss, is given by

\begin{equation}
\sum_{j=1}^n\eta{(1-\eta)^{n-j}}m = \big[1-{(1-\eta)^n}\big]\,m \,.
\end{equation}

\noindent
Figure 4 shows the quantity $[1-{(1-\eta)^n}]$, the fraction of the mass of the first target $(m)$ that is lost
after $n$ collisions. To be conservative, we show this quantity for three small values of $\eta=5\%, 10\%, 15\%$. 
As shown here, although the amount of the mass lost in one impact is small, after only a few impacts, 
the total contribution of the first term of equation (1) becomes so large than it cannot be ignored
(recall that in our perfect-merging simulations, the total number of embryo-embryo collisions that resulted
in the formation of a planet was between 6 and 11).
Applying similar analysis to other terms of equation (1), combined with the fact that planetary embryos, in addition
to colliding with other embryos, are also impacted by planetesimals will clearly demonstrate that the error in 
ignoring mass-loss in the course of the formation of a planet is not negligible. 
In the following, we demonstrate the latter for the final planets of figure 1 by directly calculating 
the amount of mass that is lost in each collision during their formation.

\subsection{Mass-loss in Figure 1}

To determine the amount of the mass that is lost during the accretion of the final three planets of figure 1, 
it is necessary to identify the seed embryo for each planet. Because in a perfect-merging collision,
the entire mass of the impactor is accreted by the target, it is possible to identify the seed object 
by tracking back through collisions. Table 2 shows this for the final planets of figure 1. To identify the planets,
we have labeled them by numbers 1, 2 and 3 corresponding to the planet on the left, middle and right in the last
panel of figure 1, respectively. We have also given their final semimajor axes to further ensure their identification.

Table 2 also shows the initial mass and water content of each seed embryo and the number of its collisions.
As shown here, in addition to impacting by embryos (shown by the entry ``Collisions with embryos''),
each growing seed was also impacted by a large number of planetesimals (shown by ``Collisions with planetesimals'').
It is important to note that during the evolution of the system, some of the embryos that collided with a growing 
seed, themselves were impacted by other embryos and planetesimals. As a result, the final
mass of a planet, and with the same token, the total mass of its collisional debris, are the results of a larger
number of impacts. We have shown these quantities by
``Total embryo-embryo collisions'' and ``Total embryo-planetesimal collisions'' for each seed. 

To calculate the final mass of a seed and the amount of the mass that was lost during its growth, 
we assumed that all embryo-planetesimal collisions were perfectly inelastic (produced no debris) and 
resulted in the complete accretion of the planetesimal. Therefore, after each collision of a planetesimal
with an embryo, we added the mass of the planetesimal (20\% lunar-mass) to the mass of that embryo.

To calculate the collisional debris produced in each embryo-embryo impact, we used the results shown in
figure 3. As indicated by this figure, for a large portion of the parameter-space, the amount of the mass
lost in each impact is at a level smaller than 20\%. We, therefore, took a conservative approach and assumed 
that in all embryo-embryo collisions, 10\% of the total colliding mass was lost in the form of debris or 
scattered fragments. We did not consider the re-accretion of the produced debris.

The entries ``Total mass-loss'' and ``Final mass'' in table 2 show the results.
We would like to emphasize that these entries are only for their corresponding seed embryos and do not
indicate the mass-loss for the entire simulation.
As expected, the total amount of mass that is lost in the formation of each of these planets is substantial, varying 
between 1.1 to almost 3.5 Mars-masses. These values are 53\% to 71\% smaller than their perfect-merging 
values and show $\sim 30\%$ to 45\% error when the results of perfect-merging simulations are used
as an approximation to realistic models.

\subsection{Water-loss in Figure 1}

The loss of water occurs when a hydrated embryo collides with a growing planet. Because not all embryos that
are accreted by a planet carry water, the total amount of water that is lost during the formation of a planet
will depend on the number of the hydrated embryos that collide with it,
the time of their collision, and whether these embryos themselves had experienced impacts prior to being
accreted by the planet.  For instance, Planet-1 acquired its water when its seed embryo-8 had its last 
collision with a hydrated embryo from the region of 2-2.5 AU. Recall that embryos from that region carry 
2\% water. The hydrated embryo, itself, had been impacted by a dry embryo prior to its accretion 
by embryo-8 meaning that it had lost some of its 2\% water before its last impact. The total
loss of water in this case would then be equal to the sum of the water that is lost in the impact of the 
hydrated embryo with the dry embryo, and the amount of water that it lost in its subsequent collision 
with growing Planet-1.

Planet-2 acquired its water during its last two embryo-embryo impacts. At this stage, the planet accreted 
two hydrated embryos. The first embryo did not have any prior
impacts. However, the second embryo had collided with a dry embryo before being accreted by the growing
planet. In this case, the total amount of water-loss is equal to the water that is lost during the accretion
of the first hydrated embryo, the amount of water that the second hydrated embryo lost in its collision
with a dry embryo, and the water lost during the accretion of the second hydrated embryo.
Planet-3 did not have any collision with a hydrated embryo,

Table 2 shows the fraction of the water that is lost in each of the two planets 1 and 2.
To maintain focus on demonstrating the significance of water-loss and the error in neglecting this quantity, 
we assumed that during a collision, the only material that is lost would be water. That is, when
a hydrated embryo undergoes a collision, the amount of its water decreases, however, the amount of its rock will stay 
intact. As a result, all non-hydrated embryos will maintain their full masses during a collision, and the 
only loss will be in the water of hydrated embryos. We also assumed that in any collision that involved hydrated 
embryos, 10\% of the total water was lost and would never be re-accreted.

The entry ``Final WMF'' in table 2 presents the total water-content of the final planets
after all water-losses have been taken into account. In calculating this quantity, we considered no loss of
water in embryo-planetesimal collisions, however, we added the mass of each planetesimal to the mass of the
final collision outcome. The ``Perfect-Merging WMF'' assumes that no water was lost during a collision and 
the entire water content of a hydrated embryo was distributed over the full perfect-merging mass of the final 
body. As shown here, in forming Planet-1, where water was delivered during the very last accreting impact, 
there is 16\% error in using the perfect-merging scenario. This value increases to over 18\% for Planet-2 where 
water was delivered during the last two collisions. These results demonstrate that not only does perfect-merging 
overestimate the water-content of the final bodies, the error introduced into calculations, especially for
objects with substantial amount of water is non-negligible.

\subsection{Comparison with previous studies}

In this section, we compare the production of the debris in our simulations with those of previous studies.  
Debris production during terrestrial planet formation has been studied by a few authors,
albeit in different contexts. For instance, in their study of the diversity of the outcome of giant impacts, 
\citet{Stewart12} investigated the effect of the collision model of \citet{Leinhardt12} on the late stage of 
terrestrial planet formation by carrying out Monte Carlo simulations of the formation of a single planet.
These authors stated that the mass of the debris produced in their single-planet formation simulations was on 
average $\sim 15\%$ of the mass of the final planet. Taking at face value and applying to the result of figure 1, 
this percentage suggests approximately 1.2 Mars-mass of debris for planets 1 and 2, and 0.5 Mars-mass of debris 
for planet 3, which are almost half the ``Total mass-loss'' shown in table 2. As we explain below,
such a low amount of debris is unrealistic and has roots in the fact that they were obtained from Monte Carlo 
simulations rather than actual N-body integrations. Unlike the latter, the final results of Monte Carlo 
simulations are not the product of the natural evolution of the initial system, and do not include the mutual 
interactions of all bodies. Also, the extreme sensitivity of N-body simulations to the initial conditions,
combined with the fact that their final results are also extremely sensitive to the type of the codes
in which they have been written, the type of the compiler, and the internal round-off errors associated 
with the architecture of the computers on which they have been carried out does not allow for 
randomly selecting initial conditions from different simulations. While Monte Carlo results are informative, 
as \citet{Stewart12} have also mentioned, they cannot replace the outcome of N-body simulations. With the
same token, these results cannot be generalized to all planet formation models as well.

A realistic study of debris production requires calculating the debris for each impact, using the mass and 
orbital elements of the colliding bodies at the moment of collision, while the N-body simulation is in progress.
\citet{Genda15} carried out somewhat similar simulations. They used the results of the SPH simulations of 
giant impacts by \citet{Genda12} and determined the total mass of collisional fragments during embryo-embryo collisions 
in the N-body simulations of \citet{Kokubo10}. Among many of their findings, these authors reported that an 
average of 4.2 Mars-masses (corresponding to $\sim 18\%$ of the initial mass of each system) turned into fragments 
in all their 50 N-body simulations. They also found that on average, in each impact, an equivalent of 0.2 Mars-mass was 
converted into debris. 

\citet{Genda15} stated that their 18\% value is consistent with the 15\% debris production reported
by \citet{Stewart12}. However, as we explain below, this comparison is not entirely valid. 
First, as mentioned earlier, due to their nature (i.e., obtained from Monte Carlo simulations), results 
by \citet{Stewart12} are not generalizable and cannot be used in lieu of the results of N-body integrations. 
In fact, the 18\% fragment mass reported by \citet{Genda15} is more applicable to
general N-body simulations than the results by \citet{Stewart12}. Second, it is important to note that 
the 15\% value reported by \citet{Stewart12}
is 15\% of the mass of the {\it final planet} whereas the 18\% value reported by \citet{Genda15} is 18\% of the 
{\it initial mass} of the system (in all their N-body simulations, the initial mass was 2.3 Earth-masses). 
The two percentages have been obtained using different methods and cannot be compared. 

Interestingly,
if the 18\% value reported by \citet{Genda15} is applied to the simulation of figure 1 where the initial mass of the
system is $\sim 5$ Earth-masses, the mass of the debris will be $\sim 9$ Mars-masses which is only slightly larger than
the sum of all three values of ``Total mass-loss'' reported in table 2 (2.52 + 3.46 + 1.10 = 7.08 Mars-masses).
Also, as reported by \citet{Genda15}, each of their giant impacts produced fragments with the total mass of 0.2 Mars-masses.
Multiplying this value by the total number of embryo-embryo collisions in table 2, the amount of mass-loss
will range between 1.8 to 3.6 Mars-masses, consistent with the total mass-loss found in our simulation.

It is, however, important to mention that the above 18\% fragment-mass and 0.2 Mars-mass debris per impact 
must be taken with caution. The reason is that when calculating the debris in a giant impact, \citet{Genda15}
did not take into account the loss of mass in previous impacts of the same body. For instance, if an object X was impacted
by a body A and later by another body B, the loss of mass on XA due to the X-A impact was not considered when the
debris produced in XA-B impact was calculated (instead, the full perfect-merging mass of XA was used). 
In other words, and as also mentioned by \citet{Genda15}, the above 18\%
fraction and 0.2 Mars-mass have been overestimated. It is for that reason that when these values 
are applied to the total number of embryo-embryo collisions in our simulations, they result in
more mass-loss than reported in table 2. This suggests that even though all collisions do not produce the 
same amount of debris, our assumption of a fixed 10\% mass-loss per giant impacts
is conservative enough to validly demonstrate the point of our argument.

The most recent calculation of giant impact debris is that of \citet{Crespi21}. These authors used the 
outcomes of 880 SPH simulations of the collisions of protoplanetary bodies and calculated the mass of
the debris in 1356 giant impacts obtained from 11 N-body simulations of the late stage of terrestrial 
planet formation. \citet{Crespi21} reported that the least amount of the debris 
produced in each of their embryo-embryo impacts was approximately 0.42 percent of the total colliding masses. 
These authors acknowledged that this value is too conservative and that more realistically, 18\% to 24\% of the 
initial mass of a system may be converted into debris.

The above 18-24\% range is obtained from the results of the current state-of-the-art simulations of terrestrial
planet formation by \citet{Burger20}. As mentioned earlier, accurate calculations of collisional mass-loss 
requires determining the amount of the debris for each impact, at the moment that it occurs, and by including the post-collision 
bodies with their post-collision masses in the N-body integrations, without interrupting the process.
\citet{Burger20} carried out such simulations. These authors introduced, for the first time, a hybrid approach to N-body 
simulations in which each giant impact is resolved accurately with an SPH code and its results are inserted 
back into the N-body integrations while the latter is in progress. \citet{Burger20} calculated the total amount 
of the collisional debris for each N-body simulation and showed that it varies between
1.1 and 1.5 Earth-masses (see the entry ``$M_{\rm col-losses}$'' in their table 2). These authors stated that
the initial mass of each of their systems was 6.1 Earth-masses indicating that 18-24\% of the
initial material was lost as collisional debris. Applying this range of mass-loss to our simulations, where initially
each system had $\sim 5$ Earth-masses of material, the total mass lost during the formation of the
final system of figure 1 would be at least at the level of 9 Mars-masses which is slightly larger than the 
7.08 Mars-masses reported in table 2. Once again, this agreement with previous results indicates that our 
10\% loss per impact is not far from reality, and the results reported here are reliable.

\section{Discussion and Concluding Remarks}

Motivated by the argument that the differences between the results of the perfect-merging simulations of
terrestrial planet formation and those in which collisions are simulated more accurately are so small 
that perfect-merging results can be used as a quantitatively acceptable approximation to realistic models,
we studied the effect of the loss of material during giant impacts 
on the mass and water content of the final planets to examine the validity of the above argument. 
We carried out an analytic study and showed that when
the formation of a planet requires more than one or two giant impacts (which, most often is the case in actual N-body
simulations), the cumulative loss of mass during collisions will be substantially large and cannot be ignored (figure 4).

To demonstrate the latter, we carried out a large number of SPH simulations of 
collisions between two Ceres-sized embryos and determined the amount of the mass and water that was lost in each collision.
We applied these results to embryo-embryo collisions in a sample simulation of the classical model
of terrestrial planet formation and showed that even when the amount of mass-loss in a single 
collision is small, the cumulative amount of the mass that is lost in all collisions that result in the formation
of a planet are so large that if perfect-merging approach is used, considerable error will be introduced 
to the final results. We also showed that not only does perfect-merging highly overestimate 
the mass of the final planets (at times by several folds, see table 2), it also overestimates the water-contents 
of the final bodies giving the false impression that some of the terrestrial planets in our solar system might 
have carried a large amount of water in the past. 

To maintain focus on demonstrating the significance of the mass and water losses during collisions,
we made two simplifying assumptions. First we assumed that irrespective of the size, material composition, 
and impact parameters of an embryo, same percentage of the total mass or total water of colliding 
bodies would be lost in each embryo-embryo collision. This mass-loss can be in the form of debris,
scattered fragments, and evaporated water. 
Because from our SPH simulations of the collisions of Ceres-sized bodies (figure 3), 
for a large range of the parameters, the amount of loss is at the level of 20\%, we took a 
conservative approach and assumed that in each collision, 10\% of the total mass or water is lost in the form 
of debris. As demonstrated in Section 3.2, these assumptions resulted in 1.10 to 3.46 
Mars-masses of debris in the formation of each planet of figure 1 indicating large errors in using perfect-merging 
results as an approximation. A comparison with previous studies confirmed that although the assumption of
a constant, fixed mass-loss per impact is not fully realistic, our 10\%  assumption is an appropriate 
approximation to reliably demonstrate the point of our argument, and the results produced with this assumption 
can be taken confidently.

That the cumulative collisional mass-loss during the formation of a planet is typically so large that it does not allow
perfect-merging to be a valid approximation to realistic models is based on the argument that
if these mass-losses are taken into account (that is, N-body simulations are carried out while masses are
removed after each impact), the final orbital elements and, therefore, the orbital architecture
of the final planetary system will be so different from that of perfect-merging results that the latter cannot be 
considered as a viable approximation. We did not carry out such simulations (reducing the mass of colliding bodies 
by 10\% and continuing the N-body integrations). However, such simulations have in fact been carried out by 
\citet{Burger20}. In their state-of-the-art approach to simulating terrestrial planet formation where 
giant impacts are resolved using SPH simulations (and without using a collisional catalog or scaling law)
and are incorporated into the N-body integrations while the latter is in progress, these authors demonstrated 
that when collisional mass-losses are taken into account, the final planetary systems will be much different
from those obtained from perfect-merging simulations \citep[we refer the reader to figure 2 of][]{Burger20}.

It is understood that in some collisions, the amount of the lost mass may be smaller than the above 10\%. 
However, it is important to note that these smaller mass-losses
may be complemented by the debris that are produced during embryo-planetesimal collisions. 
To avoid computational complications, we assumed that in these collisions, planetesimals would be fully accreted 
and no debris would be produced. However, given the mass of individual planetesimals (0.2 lunar-mass), 
it is inevitable, especially at the early stages of the growth when the seed embryos are still small,
that some debris would in fact be generated. This debris too will contribute to the total mass- and
water-loss. In connection to the values reported in table 2, this means that although the values of 
the final masses and water-contents of the planets may not be exact, when all sources of mass- and water-loss 
are taken into account, the ranges of error given by table 2 are reliable.

In calculating the final mass of a planet, we did not consider the re-accretion of the debris.
Previous studies \citep[e.g.,][]{Benz07,Kobayashi10} have shown that the time of the production and 
re-accretion of collisional debris may be comparable, suggesting that the process of re-accretion could 
be efficient. For instance, \citet{Benz07} showed that in collisions that might have stripped the mantle 
of Mercury, 30\% of the generated debris was re-accreted. However, the recent study of giant impacts 
by \citet{Crespi21} indicates that most of the ejecta produced in giant impacts are closer to the 
central star than their points of impact, meaning that their re-accretion may not be as efficient 
as in the case of giant impacts with Mercury.

As shown by table 2, the errors in mass and water estimation increase by the number of collisions.
While this is intuitive and an expected result, it is important to discuss its implications especially in connection
to the water-contents of the final bodies. We remind that in the simulation of figure 1, planets 1 and 2 
received water through one and two impacts by hydrated embryos at the end of their formation. 
Some of those hydrated embryos had also been impacted by other embryos (and planetesimals) prior to their final 
accretion. Those number of collisions, although small, still introduced approximately 18\% error in estimating the 
WMF of the final planetary bodies when the water-loss per impact was 10\%.
For a planet with a higher WMF, not only more hydrated embryos would be
accreted, there is a high probability that those embryos would have more pre-accretion impacts as well.
In other words, in a perfect-merging scenario, as the WMF of bodies increase, so does the overestimation
of their water-contact, and most likely with a higher rate than those with small number of impacts. 
That means, when modeling the formation of the terrestrial planets of our solar system, unless proper 
calculations are made to determine the actual WMF of planets in a perfect-merging simulation, any comparison 
of the final water-contents of planets in such models with the current values of the water-contents of
terrestrial planets would be misleading and can result in drawing incorrect conclusions. For instance,
they may give the wrong impression that some of the terrestrial planets in the solar system had more water in 
the past, falsely motivating scientists to develop models to explain how those planets lost their water. 

In the sample perfect-merging simulation used in this study (figure 1), 
the number of big impacts received by each seed embryo is between 7 and 10.
This number is consistent with our understanding of the dynamical evolution of the protoplanetary disk
in the inner region of the solar system. For instance, Embryo-30, the seed for Planet-3, is close to the 
region of the influence of the $\nu_6$ secular resonance (at $\sim 1.8$ AU) where most objects are scattered 
out of the system \citep[we refere the reader to][for a complete discussion of the effect of secular 
resonances on terrestrial planet formation]{Haghighipour16}. As a result, this embryo receives the least number 
of impacts. In contrast, Embryos 8 and 28, being farther in and away from the $\nu_6$ resonance, receive 
more collisions. We would like to note that although Embryos 8 and 28 are close to the location of $\nu_5$
resonance at $\sim 0.7$ AU, as shown by \citet{Haghighipour16} and \citet{Levison03}, this resonance is 
not strong enough to have significant effects on the dynamics of planetary embryos in that region.  

In closing, we would like to emphasize that, as demonstrated by its successful history, the classical 
model and the perfect-merging recipe are still the most powerful tools for proving concepts and exploring 
the physical processes that are involved in the late stage of terrestrial planet formation. 
However, for the purpose of explaining the dynamical and compositional properties of Earth and other
terrestrial planets, and in order to be able to extend those models to other planetary systems,
it is imperative that collisions be simulated accurately and the effects of debris and 
fragments be taken into account \citep[e.g.,][]{Maindl14,Burger20}. In that respect,
perfect-merging cannot offer a scientifically viable approximation that can be used to 
draw quantitatively valid conclusions and make meaningful predictions.

\acknowledgments
NH would like to expresses his gratitude to the Information Technology division of the Institute for Astronomy 
at the University of Hawaii-Manoa for maintaining computational resources that were used for carrying out the
numerical simulations of this study. We are thankful to the anonymous referee for their critically reading 
our manuscript and constructive recommendations. Our thanks also go to Christoph Burger for fruitful discussions.
Support is acknowledged for NH through NASA grants 80NSSC18K0519 and 80NSSC21K1050, and NSF grant AST-2109285. 
TIM acknowledges support from Austrian Science Fund (FWF) project P33351-N.

\clearpage
\begin{figure}
\hskip -30pt
\includegraphics[scale=0.65]{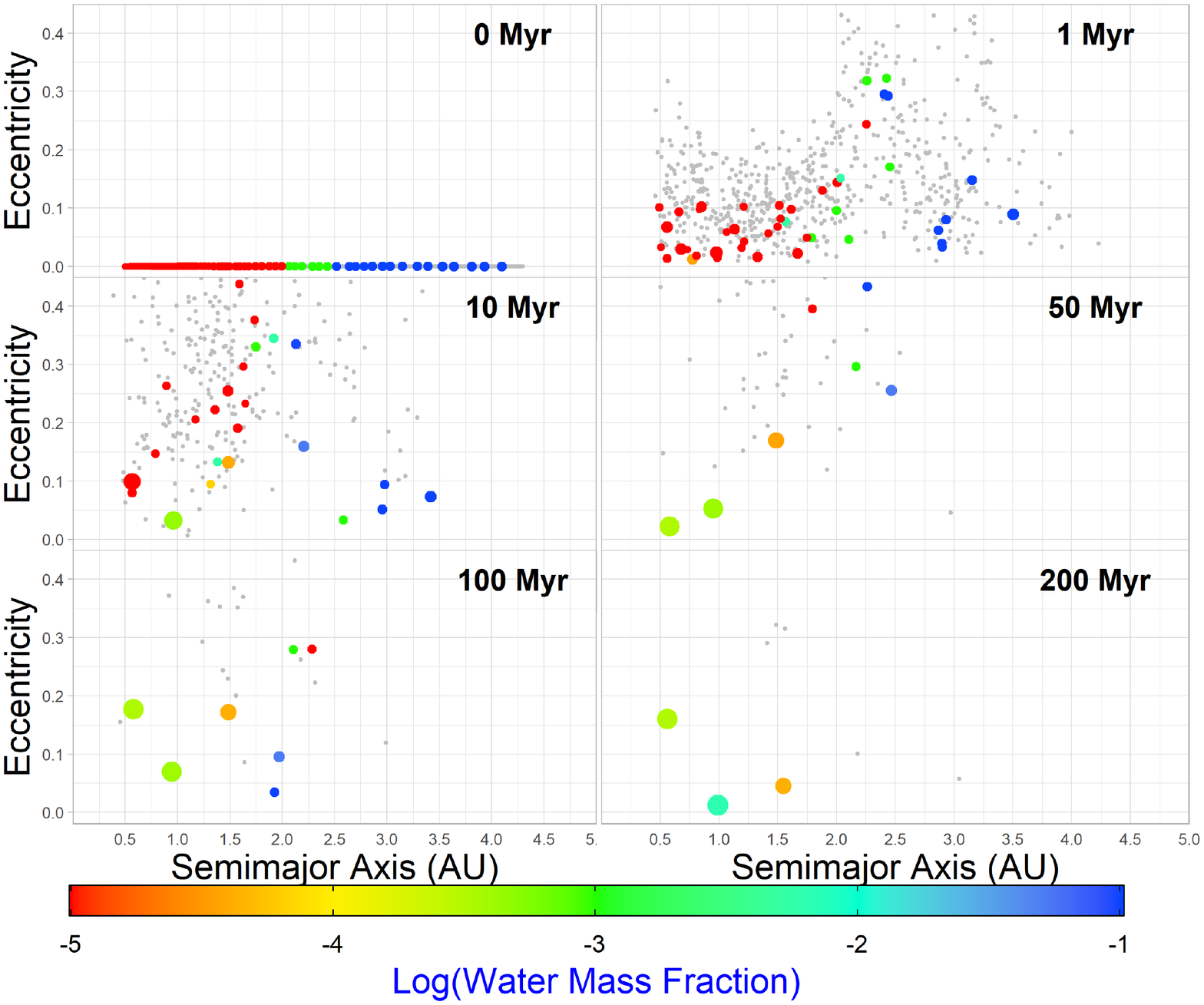}
\caption{Sample of the simulations of terrestrial planet formation in the classical model. 
In this simulation, Jupiter and Saturn are in their current orbits and carry their current
eccentricities. Planetary embryos are color coded based on their water-content. Planetesimals are
shown by color gray in the background. The masses of the embryos are proportional to their radii.}
\label{fig1}
\end{figure}

\clearpage
\begin{figure}
\vskip -6in
\hskip -1in{
\includegraphics{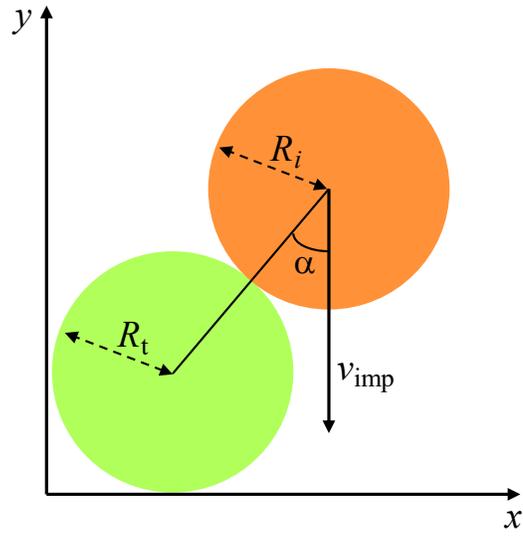}}
\vskip -4in
\caption{Geometry of an impact. The target in light green is at rest. The impactor is in orange.
The quantities $v_{\rm {imp}}$ and $\alpha$ are the impact velocity and impact angle, respectively.}
\label{fig2}
\end{figure}

\clearpage
\begin{figure}
\hskip -30pt
\includegraphics[scale=0.65]{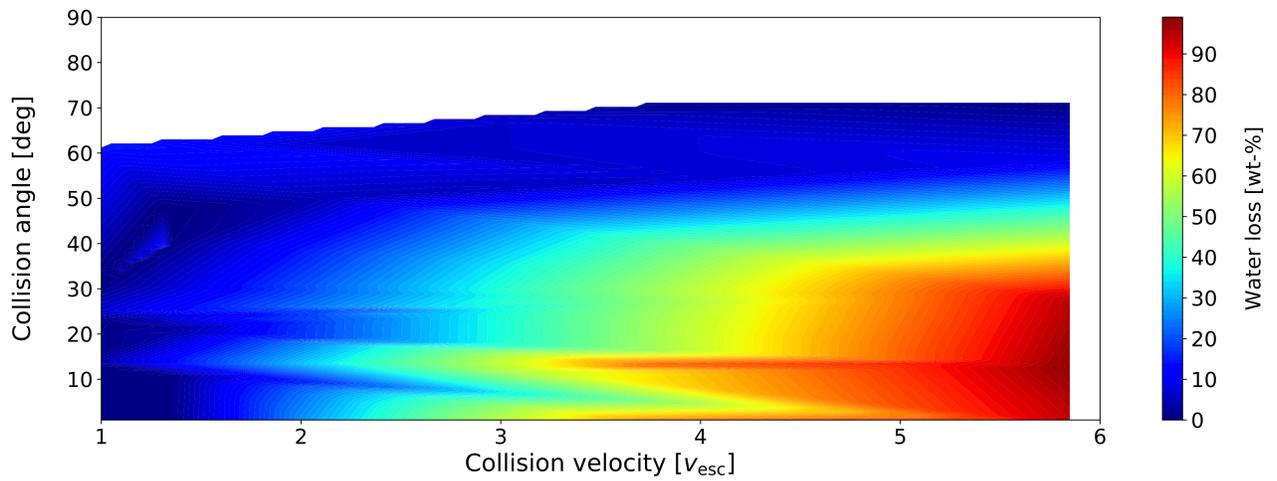}
\vskip -110pt
\caption{SPH simulation of water-loss in the collision of two Ceres-mass objects. The target is silicate 
rock (basalt) with 30\% water-mass fraction. The projectile is 100\% basaltic rock. The color-coding
represents the amount of water lost in one collision in terms of the initial water-mass.
For instance, color yellow indicates that 60\% of the total water that initally existed before the
impact was lost in the form of debris. See figure 2 for the geometry of an impact.}
\label{fig3}
\end{figure}

\clearpage
\begin{figure}
\center
\includegraphics[scale=1.5]{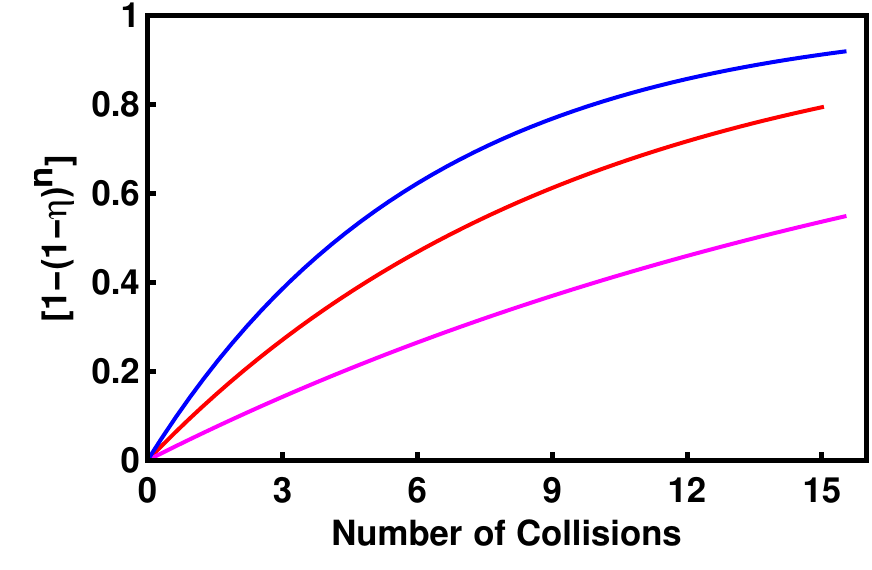}
\caption{Graph of the quantity $[1-{(1-\eta)^n}]$ in equation (2) for different number of impacts $(n)$ and
for the values of $\eta=5\%$ (Cyan), 10\% (red) and 15\% (blue). The figure shows the fraction of the mass
that is lost from the first target after $n$ collisions. As shown here, after a few impacts, the loss of
mass is not negligible.}
\label{fig4}
\end{figure}

\clearpage
\begin{deluxetable}{lccc}
\tablecaption{Material parameters for basalt and ice.}
\tablehead{ \colhead{Parameters} & \colhead{Basalt} & \colhead{Ice}  &  \colhead{Reference}} 
\startdata
Bulk density $\rho_0\,[{\mathrm{kg/m^{3}}}]$ & 2700 & 917 & \citep{Melosh96} \\
$A_\mathrm{T}$ (Tillotson EOS parameter) [GPa]  &  26.7  &  9.47  & \citep{Melosh96} \\
$B_\mathrm{T}$ (Tillotson EOS parameter) [GPa]  &  26.7  &  9.47 & \citep{Melosh96} \\
$E_0$ (Tillotson EOS parameter) [$\mathrm{MJ/kg}$]  &  487  &  10 & \citep{Melosh96} \\
$E_\mathrm{iv}$ (Tillotson EOS parameter) [$\mathrm{MJ/kg}$]  &  4.72  & 0.773  & \citep{Melosh96} \\
$E_\mathrm{cv}$ (Tillotson EOS parameter) [$\mathrm{MJ/kg}$]  &  18.2  & 3.04 & \citep{Melosh96} \\
$a_\mathrm{T}$  (Tillotson EOS parameter) &  0.5  &  0.3  & \citep{Melosh96} \\
$b_\mathrm{T}$  (Tillotson EOS parameter) &  1.5  &  0.1  & \citep{Melosh96} \\
$\alpha_\mathrm{T}$ (Tillotson EOS parameter)  &  5  &  10  & \citep{Melosh96} \\
$\beta_\mathrm{T}$  (Tillotson EOS parameter) &  5  &  5  & \citep{Melosh96} \\
$K$ (Bulk modlus) [GPa]  &  36.7  &  9.47  &  \citep{Benz99} \\
$\mu$ (Shear modulus) [GPa]  &  22.7  &  2.8  &  \citep{Benz99} \\
$Y_0$ (Yield stress) [GPa]  &  3.5  &  1  &  \citep{Benz99} \\

\enddata
\end{deluxetable}

\clearpage
\begin{deluxetable}{lccc}
\tablecaption{Post-collision mass and water-content of the final planets in figure 1}
\tablehead{
\colhead{}  & \colhead{Planet-1} & \colhead{{\hskip 30pt Planet-2}} \hskip 30pt  & \colhead{Planet-3}}
\startdata 
Seed embryo                           &  Embryo-8   & Embryo-28     &  Embryo-30    \\
Initial mass [Mars-mass]              &  0.15       & 0.23          &  0.24         \\
Initial semimajor axis [AU]           &  0.59       & 0.99          &  1.04         \\
Initial WMF [\%]                      &  0.001      & 0.001         &  0.001        \\
Final semimajor axis [AU]             &  0.56       & 0.99          &  1.55         \\
Final eccentricity                    &  0.16       & 0.01          &  0.04         \\
Collisions with embryos               &  10         & 10            &  7            \\
Collisions with planetesimals         &  79         & 44            &  18           \\
Total embryo-embryo collisions        &  18         & 18            &  9            \\
Total embryo-planetesimal collisions  &  171        & 125           &  42           \\
Total mass-loss [Mars-mass]           &  2.52       & 3.46          &  1.10         \\
Final mass [Mars-mass]                &  4.53       & 4.67          &  2.27         \\
Perfect-Merging mass [Mars-mass]      &  7.31       & 8.18          &  3.17         \\
Error in mass [\%]                    &  38         & 43            &  29           \\ 
Final WMF [\%]                        &  0.08       & 0.18          &               \\
Perfect-Merging WMF [\%]              &  0.10       & 0.22          &               \\
Error in WMF [\%]                     &  16         & 18.3          &               \\
\enddata
\end{deluxetable}


\begin{thebibliography}{}

\bibitem[Agnor et al. (1999)]{Agnor99}
Agnor, C. B., Canup, R. M. \& Levison, H. F. 1999, Icarus, 142, 219

\bibitem[Agnor \& Asphaug (2004)]{Agnor04}
Agnor, C. \& Asphaug, E. 2004, ApJL, 613, L157

\bibitem[Asphaug (2009)]{Asphaug09}
Asphaug, E. 2009, AREPS, 37, 413 

\bibitem[Benz \& Asphaug (1995)]{Benz95}
Benz, W. \& Asphaug, E. 1995, Comput. Phys. Comm., 87, 253

\bibitem[Benz \& Asphaug (1999)]{Benz99}
Benz, W. \& Asphaug, E. 1999, Icarus, 142, 5

\bibitem[Benz et al. (2007)]{Benz07}
Benz, W., Anic, A., Horner, J. \& Whitby, J. A. 2007, SSRv, 132, 189

\bibitem[Bonsor et al. (2015)]{Bonsor15}
Bonsor, A., Leinhardt, Z. M., Carter, P. J., Elliott, T., Walter, M. J. \& Stewart, S. T. 2015, Icarus, 247, 291

\bibitem[Burger et al. (2018)]{Burger18}
Burger, C., Maindl, T. I. \& Sch\"afer, C. M. 2018, CeMDA, 130, 2

\bibitem[Burger et al. (2020)]{Burger20}
Burger, C., Bazs\'o, \'A. \& Sch\"afer, C. M. 2020, A\&A, 634, id.A76 

\bibitem[Carter et al. (2015)]{Carter15}
Carter, P. J., Leinhardt, Z. M., Elliott, T., Walter, M. J. \& Stewart, S. T. 2015, ApJ, 813, article id. 72

\bibitem[Carter et al. (2018)]{Carter18}
Carter, P. J., Leinhardt, Z., M., Elliott, T., Stewart, S. T. \& Walter, M. J. 2018, E\&PSL, 484, 276 

\bibitem[Chambers \& Wetherill (1998)]{Chambers98}
Chambers, J. E. \& Wetherill, G. W. 1998, Icarus, 136, 304

\bibitem[Chambers (1999)]{Chambers99}
Chambers, J. E. 1999, MNRAS, 304, 793

\bibitem[Chambers (2001)]{Chambers01a}
Chambers, J. E. 2001, Icarus, 152, 205

\bibitem[Chambers \& Wetherill (2001)]{Chambers01b}
Chambers, J. E. \& Wetherill, G. W. 2001, M\&PS, 36, 381

\bibitem[Chambers (2013)]{Chambers13}
Chambers, J. E. 2013, Icarus, 224, 43

\bibitem[Clement et al. (2019a)]{Clement19a}
Clement, M. S., Kaib, N. A., Raymond, S. N., Chambers, J., E. \& Walsh, K. J. 2019a, Icarus, 321, 778

\bibitem[Clement et al. (2019b)]{Clement19b}
Clement, M. S., Kaib, N. A. \& Chambers, J. E. 2019b, AJ, 157, article id. 208 

\bibitem[Clement et al. (2021a)]{Clement21a}
Clement, M. S., Chambers, J. E. \& Jackson, A. P. 2021a, AJ, 161, id.240

\bibitem[Clement \& Chambers (2021)]{Clement21c}
Clement, M. S. \& Chambers, J. E. 2021b, AJ, 162, id.3

\bibitem[Clement et al. (2021b)]{Clement21b}
Clement, M. S., Kaib, N. A., Raymond, S. N. \& Chambers, J., E. 2021c, Icarus, 367, article id. 114585

\bibitem[Collins et al. (2004)]{Collins04}
Collins, G. S., Melosh, H. J. \& Ivanov, B. A. 2004, M\&PS, 39, 217

\bibitem[Crespi et al. (2021)]{Crespi21}
Crespi,S., Dobbs-Dixon,I., Georgakarakos, N., Haghighipour, N., Maindl, T. I., 
Sch\"afer, C. M. \& Winter, P. M. 2021, MNRAS 508, 6013

\bibitem[Darriba \& Haghighipour (2021)]{Darriba21}
Darriba, L. A. \& Haghighipour, N. 2021, ApJ (submitted)

\bibitem[Genda et al. (2012)]{Genda12}
Genda, H., Kokubo, E. \& Ida, S. 2012, ApJ, 744, 137

\bibitem[Genda et al. (2015)]{Genda15}
Genda, H., Kobayashi, H. \&  Kokubo, E. 2015, ApJ, 810, 136

\bibitem[Grady \& Kipp (1980)]{Grady80}
Grady, D. E. \& Kipp, M. E. 1980, Int. J. Rock Mech. Min. Sci. Geomech. Abstr., 17, 147

\bibitem[Haghighipour \& Winter (2016)]{Haghighipour16}
Haghighipour, N. \& Winter O. C. 2016, CeMDA, 124, 235

\bibitem[Ida \& Makino (1992)]{Ida92}
Ida, S. \& Makino, J., 1992, Icarus, 98, 28

\bibitem[Izidoro et al. (2013)]{Izidoro13}
Izidoro, A., Torres, K. S., Winter, O. C. \& Haghighipour, N. 2013, ApJ, 767, 54 

\bibitem[Izidoro et al. (2014)]{Izidoro14}
Izidoro, A., Haghighipour, N., Winter, O. C. \& Tsuchida, M. 2014, ApJ 782, 31

\bibitem[Jutzi (2015)]{Jutzi15}
Jutzi, M. 2015, P\&SS, 107, 3

\bibitem[Kobayashi \& Tanaka (2010)]{Kobayashi10}
Kobayashi, H. \& Tanaka, H. 2010, Icarus, 206, 735

\bibitem[Kokubo \& Ida (1998)]{Kokubo98}
Kokubo, E. \& Ida, S. 1998, Icarus, 131, 171

\bibitem[Kokubo et al. (2000)]{Kokubo00}
Kokubo, E. \& Ida, S. 2010, Icarus, 143, 15

\bibitem[Kokubo et al. (2006)]{Kokubo06}
Kokubo, E., Kominami, J. \& Ida, S. 2006, ApJ, 642, 1131

\bibitem[Kokubo \& Genda (2010)]{Kokubo10}
Kokubo, E. \& Genda, H. 2010, ApJL, 714L, 21

\bibitem[Lange et al. (1984)]{Lange84}
Lange, M. A., Ahrens, T. J. \& Boslough, M. B. 1984, Icar, 58, 383

\bibitem[Leinhardt \& Stewart (2012)]{Leinhardt12}
Leinhardt, Z. M. \& Stewart, S. T. 2012, ApJ, 745, 79 

\bibitem[Levison \& Agnor (2003)]{Levison03}
Levison, H. F. \& Agnor C. 2003, AJ, 125, 2692

\bibitem[Maindl et al. (2013)]{Maindl13}
Maindl, T. I., Sch\"afer, C., Speith, R., S\"uli, \'A., Forg\'acs-Dajka, E. \& Dvorak, R. 2013, Astron. Nachr., 334, 996

\bibitem[Maindl et al. (2014)]{Maindl14}
Maindl, T. I., Dvorak, R., Sch\"afer, C. \& Speith, R. 2014, in: Complex Planetary Systems, Proceedings of the 
International Astronomical Union Symposium, 310, 138

\bibitem[Marcus et al. (2009)]{Marcus09}
Marcus, R. A., Stewart, S. T., Sasselov, D. \& Hernquist, L., 2009, ApJL, 700, L118

\bibitem[Marcus et al. (2010)]{Marcus10}
Marcus, R. A., Sasselov, D., Stewart, S. T. \& Hernquist, L. 2010, ApJL, 719, L45

\bibitem[Melosh (1996)]{Melosh96}
Melosh, H. J. 1996, Impact Cratering (Oxford: Oxford Univ. Press)

\bibitem[Melosh \& Ryan (1997)]{Melosh97}
Melosh, H. J. \& Ryan, E. V. 1997, Icar, 129, 562

\bibitem[Michel (2009)]{Michel09}
Michel, P. 2009, Lect. Notes Phys., 758, 99

\bibitem[Monaghan \& Gingold (1983)]{Monaghan83}
Monaghan, J. J. \& Gingold, R. A. 1983, J. Comput. Phys., 52, 374

\bibitem[Morbidelli et al. (2000)]{Morbidelli00}
Morbidelli, A., Chambers, J., Lunine, J. I., Petit, J. M., Robert, F., Valsecchi, G. B. \& Cyr, K. E. 2000,  
M\&PS, 35, 1309

\bibitem[Morishima et al. (2008)]{Morishima08}
Morishima, R., Schmidt, M. W., Stadel, J. \& Moore, B. 2008, ApJ, 685, 1247

\bibitem[Nakamura et al. (2007)]{Nakamura07}
Nakamura, A. M., Michel, P., \& Setoh, M. 2007, JGRE, 112, E02001

\bibitem[Nakamura \& Michel (2009)]{Nakamura09}
Nakamura, A. M. \& Michel, P. 2009, Lect. Notes Phys., 758, 71

\bibitem[Nesvorny et al. (2021)]{Nesvorny21}
Nesvorny, D., Roig, F. V. \& Deienno R. 2021, AJ, 161, 50

\bibitem[O'Brien et al. (2006)]{Obrien06}
O’Brien, D. P., Morbidelli, A. \& Levison, H. F. 2006, Icarus, 184, 39

\bibitem[O'Brien et al. (2014)]{Obrien14}
O’Brien, D. P., Walsh, K. J., Morbidelli, A., Raymond, S. N. \& Mandell, A. M. 2014, Icarus, 239, 74 

\bibitem[O'Brien et al. (2018)]{Obrien18}
O’Brien, D. P., Izidoro, A., Jacobson, S. A., Raymond, S. N., \& Rubie, D. C. 2018, Space Sci. Rev., 214, 47

\bibitem[Raymond et al (2005a)]{Raymond05a}
Raymond, S. N., Quinn, T. \& Lunine, J. I. 2005a, Icarus, 177, 256

\bibitem[Raymond et al (2005b)]{Raymond05b}
Raymond, S. N., Quinn, T. \& Lunine, J. I. 2005b, ApJ, 632, 670

\bibitem[Raymond et al (2009)]{Raymond09}
Raymond, S. N., O’Brien, D. P., Morbidelli, A. \& Kaib, N. A. 2009, Icarus, 203, 644

\bibitem[Sch\"afer et al. (2007)]{Schafer07}
Sch\"afer, C., Speith, R. \& Kley, W. 2007, A\&A, 470, 733

\bibitem[Sch\"afer et al. (2016)]{Schafer16}
Sch\"afer, C., Riecker, S., Maindl, T. I., Scherrer, S., Speith, R. \& Kley, W. 2016, A\&A, 590, A19

\bibitem[Sch\"afer et al. (2020)]{Schafer20}
Sch\"afer, C. M., Wandel, O. J., Burger, C., et al. 2020, Astron. Comput., 33, 100410  

\bibitem[Stewart \& Leinhardt (2012)]{Stewart12}
Stewart, S. T. \& Leinhardt, Z. M. 2012, ApJ, 751, 32

\bibitem[Tilloston (1962)]{Tillotson62}
Tillotson, J. H. 1962, Metallic Equations of State for Hyper-velocity Impact, 
General Atomic Rep. GA-3216 (San Diego, CA: General Atomic)

\bibitem[Wallace et al. (2017)]{Wallace17}
Wallace J., Tremaine, S. \& Chambers, J. 2017, AJ, 154, 175

\bibitem[Walsh et al. (2011)]{Walsh11}
Walsh, K. J., Morbidelli, A., Raymond, S. N., O’Brien, D. P. \& Mandell, A. M. 2011, Nature, 475, 206

\bibitem[Weibull (1939)]{Weibull39}
Weibull, W. A. 1939, Ingvetensk. Akad. Handl., 151, 5


\end{thebibliography}
\end{document}